# Generation of high concentration nanobubbles based on friction tubes

Taekeun Yoo,[1] Young-Ho Yoo,[1] Suk-Joo Byun,[1] A-Ram You,[2] Chang-Hee Park,[1] Dae-Hyun Choi,[1] Eun-Hee Jun[1]

1 Fawoo Nanotech Corporation, Bucheon-si, Gyeonggi-do, Republic of Korea
2 School of Architecture, Kumoh National Institute of Technology, Gumi, Gyeongbuk, South Korea
Correspondence: Taekeun Yoo, fawoo2@snu.ac.kr; Young-Ho Yoo, fawoonano@naver.com

**Abstract**

Nanobubble-related technologies have been confirmed to be useful in various fields such as climate change and the environment as well as water-based industries such as water purification, crops, horticulture, medical care, bio, and sterilization. However, a method of mass production in real time enough to apply nano-bubbles to the industry has not yet been developed. We explored the mechanism of nano-bubble water generation by friction between water and walls and developed a tube device applying the shape of the flow path to maximize the friction in the fluid passing through the flow path. It also describes the case of real-time and low-power mass production of nanobubbles and its technical utility. We found that the friction of nanotubes alone can easily and quickly improve the production of nanobubbles with small particle size in real time; by increasing the shearing pressure while increasing the effective friction constant value, the particle size of nanobubbles can be smaller while increasing the particle concentration.

## I. Introduction

### 1. Research Background

Nanobubble in water has gained attention because it can provide versatile and sustainable method to solve various industrial problems.[1] Nanobubbles are nano-scaled gaseous cavities (typically containing air) in solutions that have changed characteristics compared to general water.[2] Generally, nanobubbles sized less than 1 micrometer can remain in fluids for an externed duration, and larger ordinary bubbles with larger sizes rise to the surface and collapse.[3] When water without additional infused gas is subjected to frictional surface with fast speed, oxygen dissolved in molecular form in the water changes into nano-bubble particles with vortex.[4] At this time, the temperature rises and the water with nanobubble has a slippery property than water. Nanobubble-related technologies have been confirmed to be useful in various fields such as climate change and the environment as well as water-based industries such as water purification, crops, horticulture, medical care, bio, and sterilization.[5]

Several studies have proposed the bulk nanobubble generation methods. Initially, ultrasonication was proposed to generate nanobubble solution.[6] Several studies have shown that electrolysis is an effective method to generate nanobubble.[7,8] However, these methods have the disadvantage of not being able to produce a large amount of nano-bubble water because they use limited end-effectors. To produce bulk nanobubble solution without end-effectors, hydrodynamic cavitation approach has been adopted but complex systems such as Venturi have been devised.[9] It also occupies a large installation space and is difficult to use for small businesses or home use. More importantly, the microbubbles generated by cavitation are mostly in the microbubble stage, and the concentration of nanobubbles in the fluid treated by the device is minimal, failing to have the unique features of nanobubble water.

### 2. Research Objective

To solve the aforementioned problems, this study created a flow path member for generating nanobubbles, which can significantly improve the micronization quality and the ability to generate nanobubbles. We expanded the friction area with the fluid and the boundary layer area of the fluid per cross-sectional area of the flow path, which is the basis for generating nanobubbles, and eventually generated nanobubbles. It also attempted to expand the utilization field of nanobubbles based on the facile installation in a separate device requiring various types of bending.

Therefore, in a unit volume of water, the water in contact with the flow path can generate good quality nanobubbles quantitatively and qualitatively based on the following conditions: (1) the larger friction surface area, (2) the higher friction coefficient, (3) the higher flow velocity, (4) the longer friction duration, and (5) the more gas in the water.

Based on the principle of generating nanobubbles using this friction method and its friction efficiency, a tube characterized by a densely formed friction surface in the flow space was developed and tested (hereinafter referred to as "nanotube").

As for nanotubes with densely formed friction surfaces, the length of the tube can be continuously formed to more than several meters while lengthening the length of the inner perimeter of the cross-section compared to the cross-sectional area of the flow path; compared to existing production methods, it enables controlling the size and concentration of nanobubbles and dramatically increasing the production volume with a smaller and lighter size, lower cost and lower power.

In this study, basic research on the production of nanobubbles using nanotubes was conducted, and methods to improve the refinement quality and production capacity of nanobubbles.

## II. Research Method and Findings

### 1. Definition of Friction Integration and Effective Friction Constant

To determine the correlation between the size and concentration of nanobubbles generated when water passes through the nanotube, the concepts of (1) friction integration (scale factor) and (2) effective friction constant were introduced as follows. The friction integration for the cross-sectional inner perimeter length and the cross-sectional area of the flow path can be defined as follows. (Table. 1)

$$\text{Friction integration} = \frac{\text{Cross−sectional inner perimeter length (mm)}}{\text{Cross−sectional area of the flow path}(\text{mm}^2)} \quad \text{(1)}$$

$$\text{Effective friction constant} = \text{Friction integration} * \text{Nanotube length (m)} \quad \text{(2)}$$

Table. 1 Friction integration in line with nanotube cross-section

| External diameter | Internal diameter | Sectional view | Cross sectional area ($mm^2$) | Internal length (mm) | Scale factor |
|---|---|---|---|---|---|
| Ø15 | Ø12.6 | | 66 | 181.2 | 2.7 |
| Ø25 | Ø22.2 | | 244.4 | 344.5 | 1.4 |
| Ø90 | Ø80 | - | 3,389 | 2,907 | 1.17 |

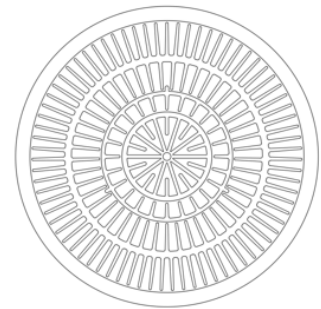

Ø90 – Sectional view

The members of the developed nanotubes are shown in Figure 1. As a result of performing several experiments described in this study, the conditions (room temperature 24°C, oxygen injection) that resulted in nanobubble particle size of around 200nm or less, and particle concentration of around 2.0e+008 Particles/ml or higher, mostly had an effective friction constant of higher than 4, and a shearing pressure of higher than 1.5 bar.

The larger the effective friction constant, the smaller the particle size and the higher the particle concentration; when the effective friction constant is lower than 4, the particle size increases and the nanobubble particle concentration decreases.

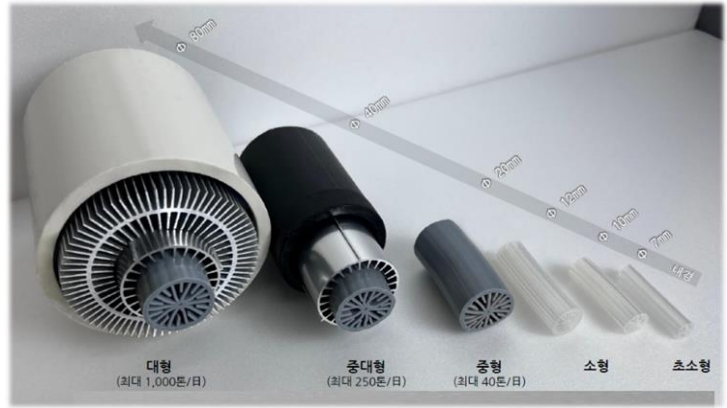

Fig. 1 Various sizes of the developed nanotubes

## 2. Nanobubble Generation via Nanotubes

### 2.1 Reagent Preparation

As for the water used to generate nanobubbles, Type 1(Ultra-pure Water) distilled water was used among the four specifications (D1193-06(2018), Standard Specification for Reagent Water) determined by the American Society for Testing and Materials (ASTM) (Table. 2)

Table 2. Standard specification for reagent water, ASTM-D1193-06(2018)

| ASTM D1193-06 | TYPE 1 | TYPE 2 | TYPE 3 | TYPE 4 |
|---|---|---|---|---|
| Conductivity, min. µS/cm (25oC) | 0.056 | 1 | 0.25 | 5 |
| Resistivity, min. MΩ-cm (25oC) | 18 | 1 | 4 | 0.2 |
| TOC, max. µg/l | 50 | 50 | 200 | No limit |
| Sodium, max. µg/l | 1 | 5 | 10 | 50 |
| Silica, max. µg/l | 3 | 3 | 500 | No limit |
| Chloride, max. µg/l | 1 | 5 | 10 | 50 |
| pH value (25oC) | – | – | – | 5.0–8.0 |

### 2.2 Configuration of Experimental Equipment

The main components of the equipment for generating nanobubbles can be classified as a sample tank, pump, and nanotube, and the specific specifications and detailed configuration are described below.

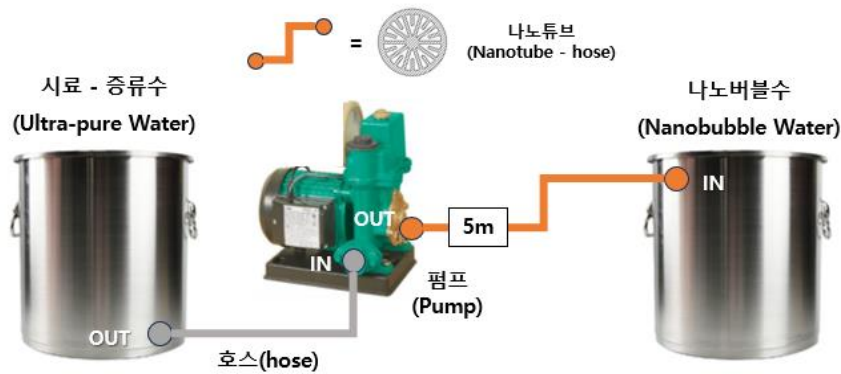

Fig. 2 Diagram of nanobubble generating equipment – for one pass use

As shown in Fig. 2, a sample tank made of SUS 304(Stainless steel, Ni 8~11%, Cr 18~20%) was filled with distilled water; a model 'PW-S354M pump for water supply' with a rated output of 350W, a pumping capacity of 26ℓ/min, and a rated power of 220V, 60Hz produced by Wilo was prepared in order to increase the flow velocity; the input end of the pump and the output end of the sample tank was connected with a hose. A nanotube with a friction integration of 1.4, an effective friction constant of 7.0, an outer diameter of Ø25, and a length of 5 meters was connected to the pump output, and the opposite end of the nanotube was connected to another sample tank made of SUS 304. After confirming the shearing pressure of the nanotube at 1.8 bar, the pump was run to generate nanobubbles without any gas input for the prepared equipment.

### 2.3 Nanobubble Particle Analysis Device: Nanosight NS300 Nanosight NS300

A particle analysis tracking device (Nanosight NS300, Malvern Panalytical. UK) with a laser with a wavelength of 532nm was utilized to confirm the particle size and concentration of the generated nanobubbles. As for the numerical results and graphs of particle analysis, the numerical results and graphs generated by Nanosight S/W after the measurement were shown as they were; to increase the reliability of the particle analysis outcomes, the raw water state was first measured before passing through the device, and then the measurement was conducted on the samples that passed through the device once and twice without gas input.

### 2.4 Experimental Results

#### 2.4.1 Measurement of Raw Water Sample

Based on the measurement of the raw water before the experiment, it was confirmed that the sample was free of foreign substances and air bubbles. (Fig. 3)

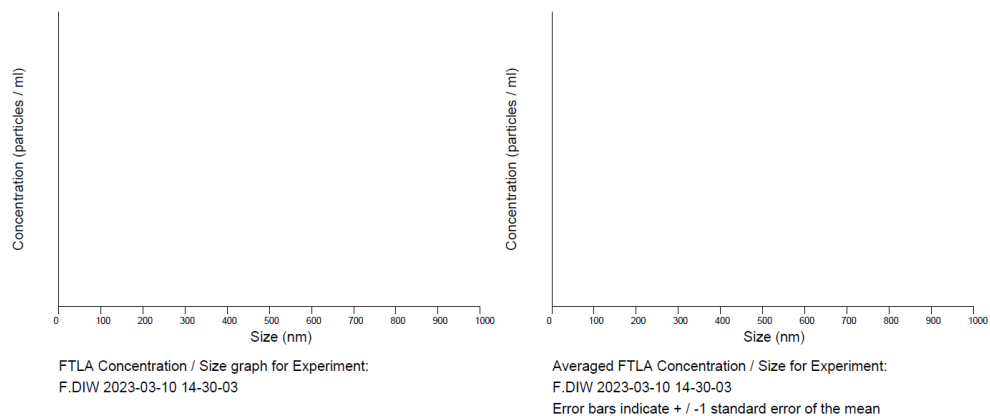

Fig. 3 Measurement of raw water samples

### 2.4.2 Measurement After One Pass

Distilled water was added to the sample tank to confirm the dissolved oxygen (DO) through the 'HI9147' measurement device by 'HANNA Co.,' confirming that the measurement result was 3.5PPM. As a result of analyzing the samples that passed through the device once without gas input five times in a row with an interval of 10 seconds, like Fig. 2-2, the average particle size of 84.8nm, and the particle concentration of 2.81e+08 Particles/ml were confirmed. (Fig. 4)

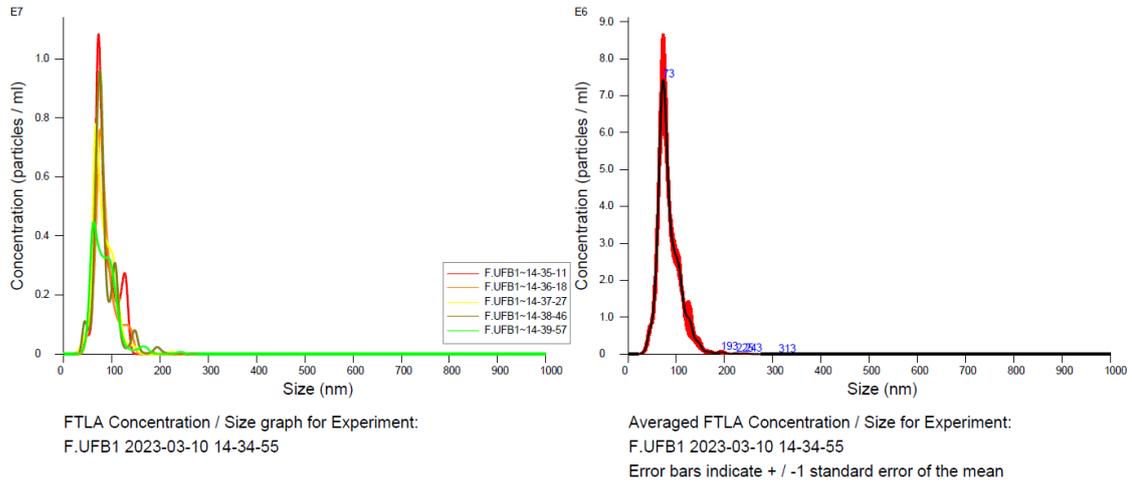

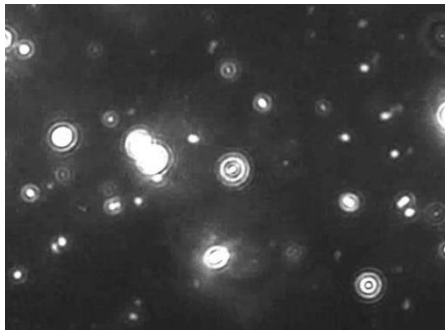

Fig. 4 Measurement after one pass through the nanobubble device (Video. 2-1)

### 2.4.3 Measurement After Two Passes

As a result of analyzing the samples that passed through the device twice without gas input five times in a row with an interval of 10 seconds, like Fig. 5, the average particle size of 84.3nm, and the particle concentration of 4.40e+08 Particles/ml were confirmed. (Fig. 6)

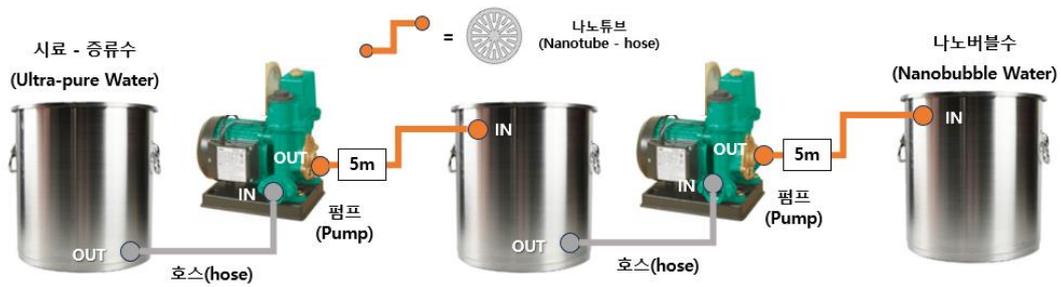

Fig. 5 Diagram of nanobubble generating equipment – for 2 passes use

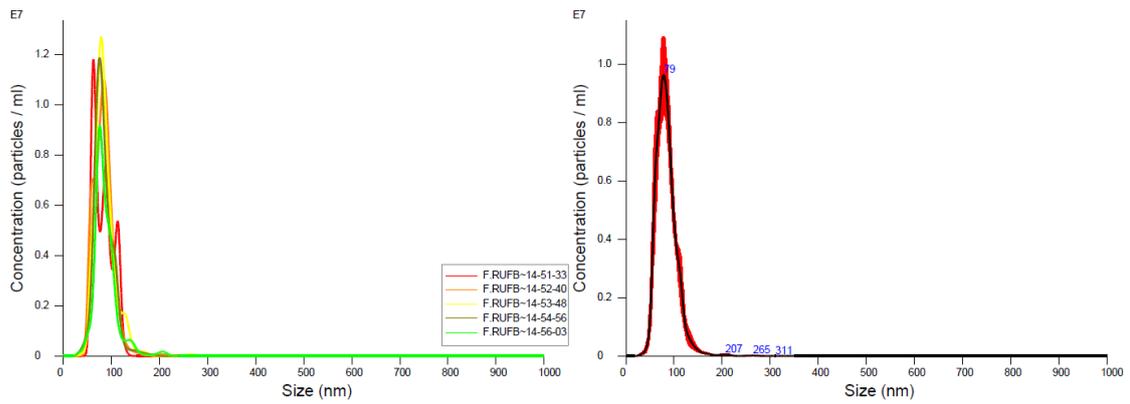

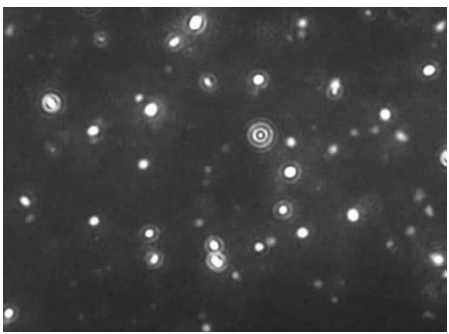

Fig. 6 Measurement after two passes through the nanobubble device (Video. 2-2)

The above experiment indicates that even without adding gas, the oxygen dissolved in water as a molecule is changed into nanobubble-size bubbles.

## 3. Nanobubble Characteristics According to Sample Temperature and Nanotube Length

### 3.1 Configuration of Experiment Equipment and Sample Preparation

The particle size and concentration were confirmed by varying the nanotube length and temperature at the same friction integration. In this experiment, tap water was used as the sample, and tap water is standardized worldwide in terms of water purification method and chemical composition, and is disinfected and sterilized by installing filtration facilities in the water supply; considering its uniform quality and facile utilization, the tap water was utilized in this experiment.

When generating the sample, the nanotube was connected to a general household sink without adding any gas; the pressure of the tap water was confirmed at 3bar, and the dissolved oxygen was confirmed to be 10ppm; the temperature of the tap was set at 24°C, and 34°C; two types of nanotube (Ø15) with lengths of 2.0m, and 3.5m were prepared, and then the nanobubble water was generated for each. As for the sample of 34°C, the temperature of the tap water that passed through the water heater was confirmed and the sample was generated. In addition, a friction coefficient of 2.7 and an effective friction constant of 5.4. which were the results of using a 2 m of Ø15 nanotubes, and an effective friction constant of 9.45, which was the result of using a 3.5 m of Ø15 nanotubes were applied.

### 3.2 Nanobubble Particle Analysis Device: Nanosight LM-10

When examining the particle size and concentration of the generated nanobubbles, a particle analysis tracking device (Nanosight LM-10, Malvern Panalytical. UK) with a laser with a wavelength of 488nm was utilized; in the case of the numerical results and graphs of the particle analysis, the numerical results and graphs generated by Nanosight S/W after the measurement were shown as they were.

## 3.3 Experimental Results

### 3.3.1　Sample Measurement after Passing through 24°C Tap Water and 2.0m Nanotubes

As a result of analyzing the sample that passed through 24°C tap water and 2.0m nanotubes five times in a row with an interval of 10 seconds, the average particle size was 137.1nm, the particle concentration was 1.99e+08 Particles/ml, and the number of particles in the measurement screen was 10.1±1 Particles/frame. (Fig. 7)

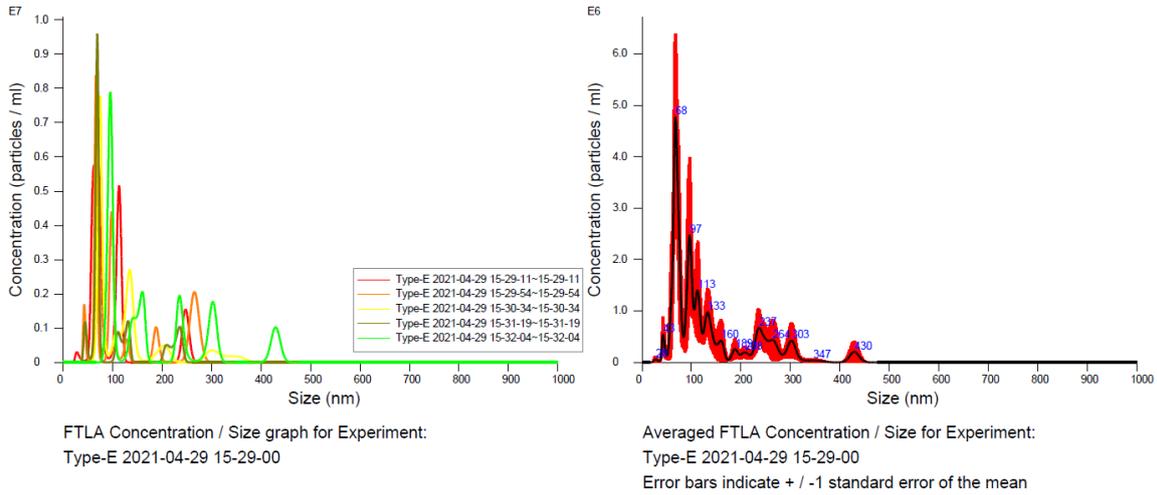

Fig. 7 Sample measurement after passing through 24°C tap water and 2.0m nanotubes

### 3.3.2　Sample Measurement after Passing through 24°C Tap Water and 3.5m Nanotubes

As a result of analyzing the sample that passed through 24°C tap water and 3.5m nanotubes five times in a row with an interval of 10 seconds, the average particle size was 133.8nm, the particle concentration was 2.09e+08 Particles/ml, and the number of particles in the measurement screen was 10.6±0.8 Particles/frame. (Fig. 8)

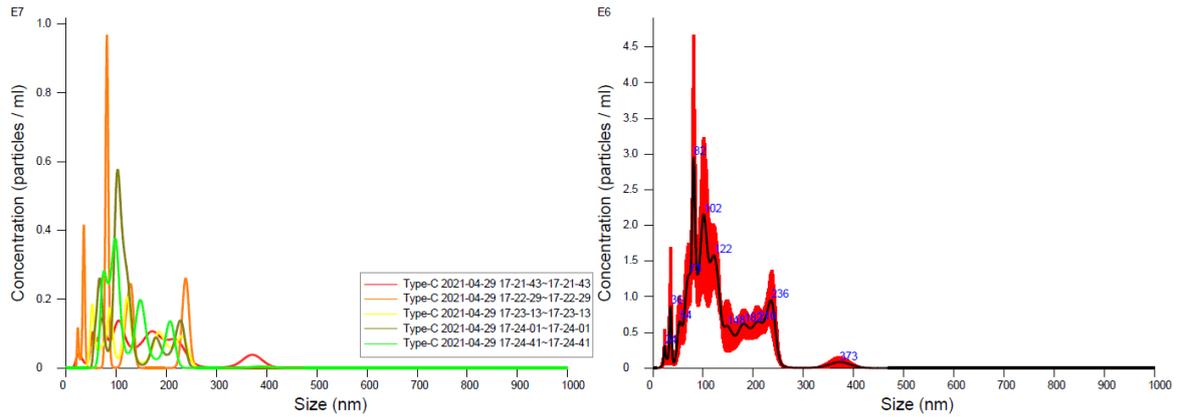

Fig. 8 Sample measurement after passing through 24°C tap water and 3.5m nanotubes

### 3.3.3 Sample Measurement after Passing through 34°C Tap Water and 2.0m Nanotubes

As a result of analyzing the sample that passed through 34°C tap water and 2.0m nanotubes five times in a row with an interval of 10 seconds, the average particle size was 104.1nm, the particle concentration was 4.40e+08 Particles/ml, and the number of particles in the measurement screen was 22.4±1.8 Particles/frame. (Fig. 9)

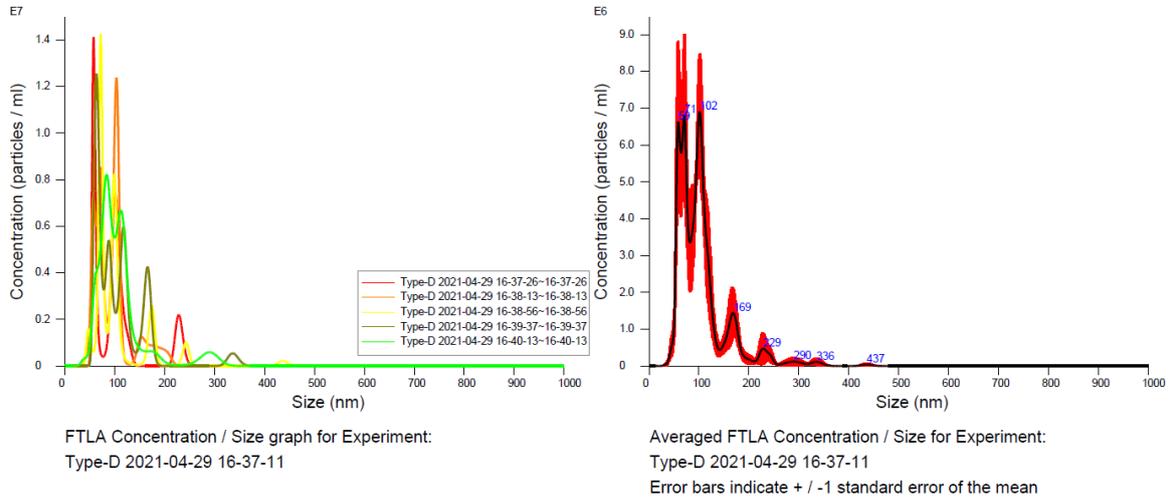

Fig. 9 Sample measurement after passing through 34°C tap water and 2.0m nanotubes

### 3.3.4 Sample Measurement after Passing through 34°C Tap Water and 3.5m Nanotubes

As a result of analyzing the sample that passed through 34°C tap water and 3.5m nanotubes five times in a row with an interval of 10 seconds, the average particle size was 108.6nm, the particle concentration was 3.14e+08 Particles/ml, and the number of particles in the measurement screen was 16±2.3 Particles/frame. (Fig. 10)

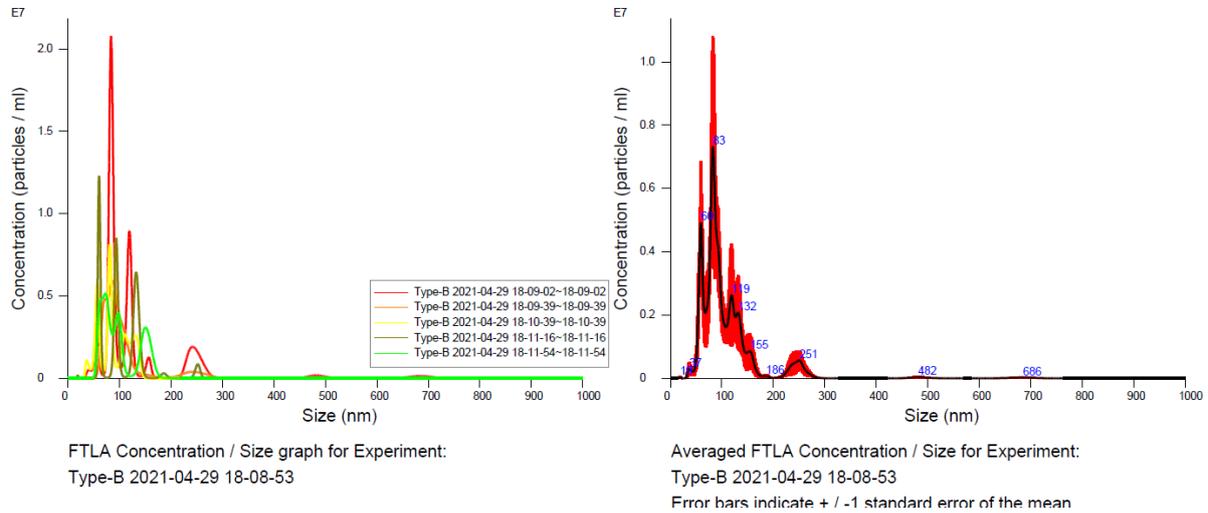

Fig. 10 Sample measurement after passing through 34°C tap water and 3.5m nanotubes

Table 3. Concentration and average size of particles in line with nanotube length and temperature

| Effective friction constant | Tube length(m) | Temperature(°C) | Concentration Particles/ml | Mean(nm) |
|---|---|---|---|---|
| **2.7** (Ø15) | 2.0m (Fig. 2-6) | 24°C | 1.99e+08 | 137.1nm |
| | 3.5m (Fig. 2-7) | | 2.09e+08 | 134.4nm |
| | 3.5m (Fig. 2-8) | 34°C | 4.40e+08 | 104.1nm |
| | 3.5m (Fig. 2-9) | | 3.14e+08 | 108.6nm |

Table 3 shows the concentration and average size of particles in line with nanotube length and temperature. The concentration and the average size of particles were compared when the nanotube length and temperature were varied under the condition of the same pressure and friction integration, as seen above. In this experiment, any gas was not input when generating the sample, but only the temperature for the same nanotube length was increased, resulting in a smaller average particle size and higher particle concentration.

Most gases increase their solubility at lower temperatures and higher pressures, but in the above experiment, the sample was created under the condition that the temperature was increased from 24℃ to 34℃ while the pressure was fixed at 3bar; it is assumed that the oxygen dissolved in the existing water became less soluble and more fluid due to the higher temperature; the flow velocity increased, and the bubbles trying to get out of the water became thinner and longer due to friction with the inner wall of the nanotube and the flow velocity; the bubbles were differentiated into small bubbles and refined, resulting in smaller average particle size and a higher particle concentration.

When comparing the particle concentration of 2m and 3.5m nanotubes at the same 34℃ temperature, the results were as follows: 4.40e+08 Particles/ml for the 2m nanotube, and 3.14e+08 Particles/ml for the 3.5m nanotube. This indicates that the friction duration may be relatively long as the nanotube length increases, but the particle concentration is relatively low because the flow velocity drops.

### 4. The Method of Increasing the Concentration of Nanobubble Particles

#### 4.1 Configuration of Experiment Equipment and Sample Preparation

To increase the concentration of nanobubble particles, a device using two pumps was constructed to conduct the experiment. As shown in Fig. 11, a sample tank made of SUS 304 was filled with tap water; a model 'a PW-S354M pump for water supply' (No. 1) with a rated output of 350W, a pumping capacity of 26ℓ/min, and a rated power of 220V, 60Hz produced by Wilo was prepared in order to increase the flow velocity; and the input end of the pump and the output end of the sample tank were connected with a hose. Then, a two-hole distribution port made of SUS304 was connected to the output of No.1 pump; two silicon nanotubes with a friction integration of 2.7, an effective friction constant of 3.24, an outer diameter of Ø15, and a length of 1.2 meters were connected in parallel; a T-type nipple was used to connect to the input of the PW-S354M pump for water supply (No. 2)' produced by Wilo with the same specifications.

Again, a two-hole distribution port made of SUS 304 was connected to the output of the No. 2 pump, and two silicon nanotubes with a friction integration of 2.7, an effective friction constant of 6.75, an outer diameter of Ø15, and a length of 2.5 meters were connected in parallel. Finally, each 2.5-meter nanotube was connected to a T-type nipple, and the sample was generated through a single row of silicon nanotubes with a friction integration of 2.7, an effective friction constant of 13.5, an outer diameter of Ø15, and a length of 5 meters (Hereinafter referred to as "special type nanobubble generator."). The shearing pressure of the 1.2-meter nanotube connected to the No.1 Pump was 4.0 bar, and the shearing pressure of the 7.5-meter (2.5+5-meter) nanotube connected to the No.2 Pump was 7.0 bar (Total nanotube length was 8.7m, and the effective friction constant was 23.49).

After confirming that the tap water temperature was 22℃ at the time of sample generation, oxygen was supplied at 2.0LPM using an oxygen gas cylinder (bombe) at the inlet of the No.1 pump of the prepared device.

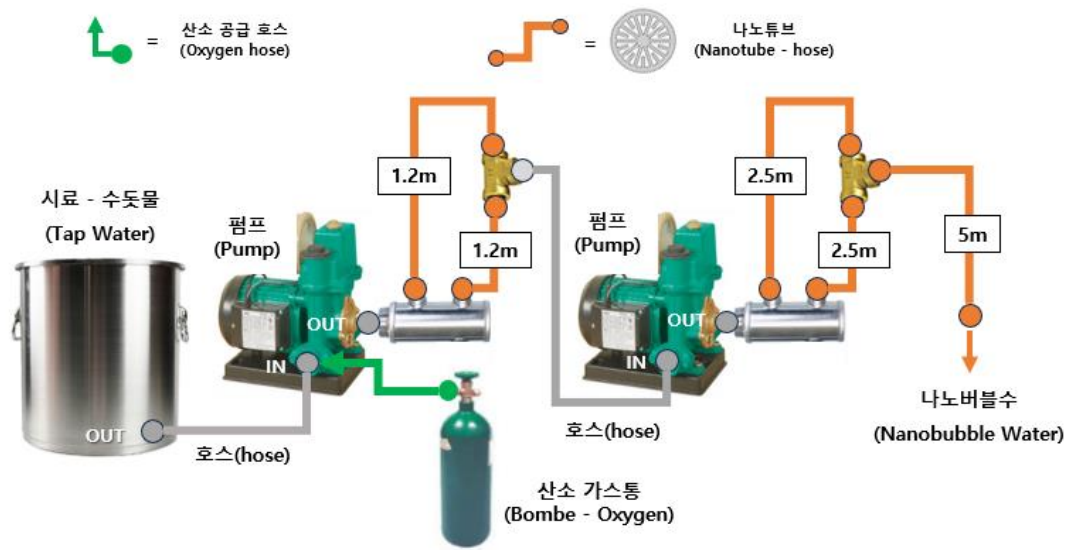

Fig. 11 Diagram of the special type nanobubble generator

## 4.2 Experimental Results

### 4.2.1 Sample Measurement after Passing Through Friction Integration of 2.7 and 8.7 meters of Nanotubes – Dilution (1:10)

As for the size and concentration of the generated nanobubble particles, the same particle analysis tracking device used in Section 3.2 was employed; as for the collected sample through the device explained in Fig. 2-10, Type 1(Ultra-pure Water) distilled water in Table 2-2 was diluted to 1:10, and analyzed five times in a row at 10-second intervals; as a result, the average particle size was 149.5nm, the number of particles in the measurement screen was 21.7±0.5 Particles/frame, and the particle concentration was 4.2e+08 Particles/ml; it can be predicted that the concentration of raw water according to the measurement results and the dilution ratio is 46.2e+08 Particles/ml (Fig. 12).

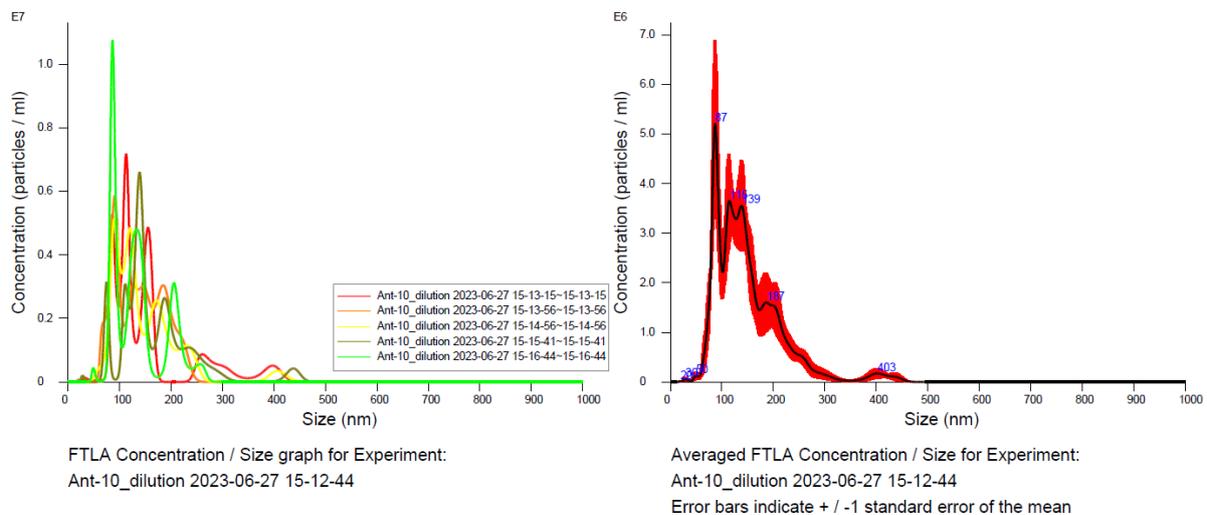

Fig. 12 Sample measurement after passing through friction integration of 2.7 and 8.7 meters of nanotubes (dilution ratio - 1:10)

### 4.2.2 Sample Measurement after Passing Through Friction Integration of 2.7 and 8.7 meters of Nanotubes - No Dilution

The undiluted raw water was analyzed five times in a low at 10-second intervals, resulting in the average particle size of 158.9nm, the number of particles in the measurement screen of 14.5±1.3 Particles/frame, and the particle concentration of 2.85e+08 Particles/ml.

The reason for the relatively low particle concentration compared to the diluted sample is speculated to be attributable to the high density of nanobubbles in the sample storage container during the 6 hours of transportation from the sample collection to the measurement equipment, where they collided with each other and became micro and disappeared (Fig. 13)

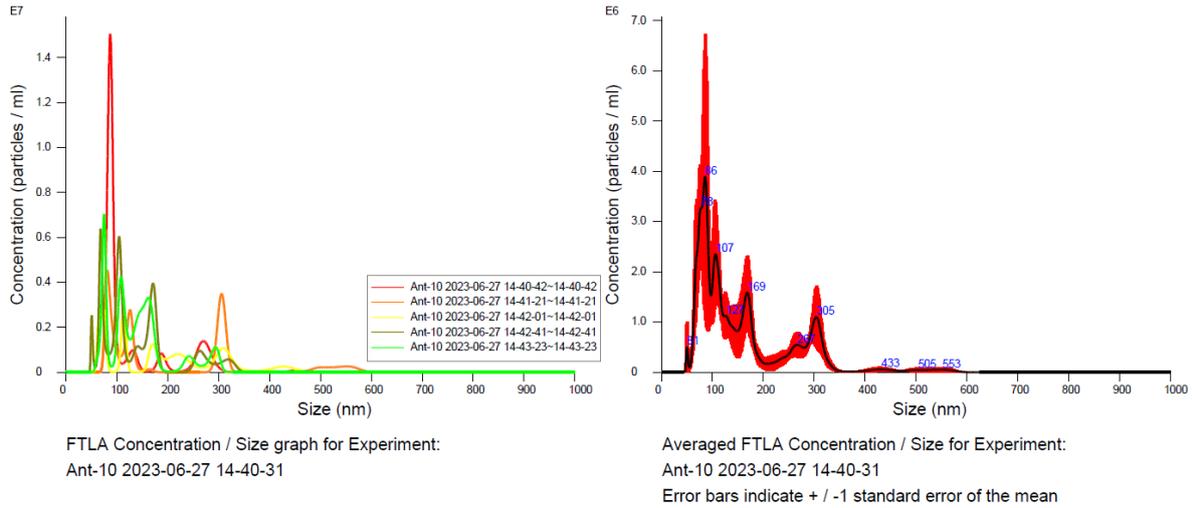

Fig. 13 Sample measurement after passing through friction integration of 2.7 and 8.7 meters of nanotubes (no dilution)

In this experiment, it is determined that the particle concentration was high by increasing the pressure with two pumps to increase the flow velocity, and connecting the nanotubes in parallel to provide sufficient flow, while having a long friction time through the high effective friction constant. By adjusting the nanotube length and applying the pump configuration and arrangement, it is assumed to be possible to generate better nanobubbles in terms of quantity and quality.

## 5. Nanobubble Characteristics According to Nanotube Length

### 5.1 Configuration of Experiment Equipment and Sample Preparation

The size and concentration of particles were confirmed by varying the nanotube length at the same friction integration. As shown in Fig. 14, a sample tank made of SUS 304 was filled with tap water; a model 'a PW-S354M pump for water supply' with a rated output of 350W, a pumping capacity of 26ℓ/min, and a rated power of 220V, 60Hz produced by Wilo was prepared in order to increase the flow velocity while confirming temperature of 26℃; and the input end of the pump and the output end of the sample tank were connected with a hose.

Six nanotubes of 3.2m, 2.9m, 2.6m, 2.3m, 1.7m, and 1.4m with a friction integration of 2.7, and an outer diameter of Ø15 were prepared and connected to the pump output end; nanobubble water was generated by supplying oxygen at 2.0 LPM using an oxygen gas cylinder (bombe) at the pump inlet of the prepared device when generating samples by length.

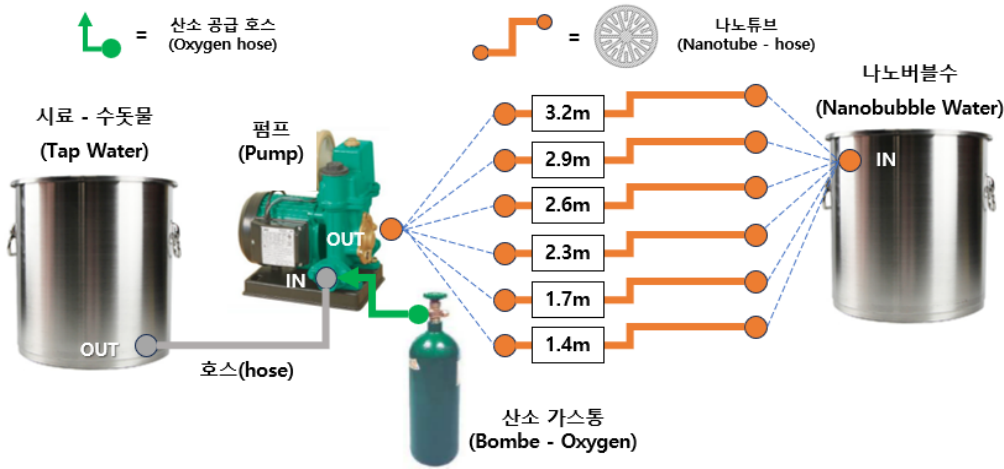

Fig. 14 Diagram of the nanobubble generator in line with the nanotube length

### 5.2 Experimental Results

As for the size and concentration of the generated nanobubble particles, the same particle analysis tracking device used in Section 3.2 was utilized; to increase the reliability of the particle size analysis results, the same length sample was analyzed three times in a row with an interval of 10 seconds.

The particle size distribution and concentration of the passed samples per nanotube length are summarized in the table below. As a result of the measurement, there are the following outcomes; (1) the average particle size of 3.2m nanotubes was 109.6nm, the concentration of particles was 4.21e+08 Particles/ml, and the number of particles in the measurement screen was 21.3 Particles/frame; (2) the average particle size of 2.9m nanotubes was 122.1nm, the concentration of particles was 2.89e+08 Particles/ml, and the number of particles in the measurement screen was 14.8 Particles/frame; (3) the average particle size of 2.6 m nanotubes was 125.5nm, the concentration of particles was 2.67e+08 Particles/ml, and the number of particles in the measurement screen was 13.6 Particles/frame, (4) the average particle size of 2.3 m nanotubes was 130.4nm, the concentration of particles was 1.80e+08 Particles/ml, and the number of particles in the measurement screen was10.2 Particles/frame, (5) the average particle size of 1.7 m nanotubes was 127.8nm, the concentration of particles was 2.34e+08 Particles/ml, and the number of particles in the measurement screen was 11.9 Particles/frame, (6) the average particle size of 1.4 m nanotubes was 135.8nm, the concentration of particles was 2.11e+08 Particles/ml, and the number of particles in the measurement screen was 10.7 Particles/frame. (Table 4)

Table 4. Size distribution and concentration measurement results of particles per nanotube length

| Tube length(m) | 3.2m | 2.9m | 2.6m | 2.3m | 1.7m | 1.4m |
|---|---|---|---|---|---|---|
| Mean (nm) | 105.4 | 106 | 130.7 | 119.3 | 133.7 | 131 |
|  | 124.4 | 135.4 | 131.2 | 128.3 | 119.6 | 144.4 |
|  | 99 | 125 | 114.5 | 143.7 | 130.2 | 131.9 |
| Average - Mean | 109.6nm | 122.1nm | 125.5nm | 130.4nm | 127.8nm | 135.8nm |
| Mode (nm) | 74.7 | 68 | 72.4 | 113.1 | 166.9 | 105.2 |
|  | 79.4 | 84.5 | 102 | 101.7 | 120.2 | 65.2 |
|  | 88.6 | 100.2 | 87.8 | 85.9 | 137.2 | 58.4 |
| Average - Mode | 80.9nm | 84.2nm | 87.4nm | 100.2nm | 141.4nm | 76.3nm |
| Concentration (paticles/ml) | 3.55E+08 | 2.51E+08 | 2.63E+08 | 2.23E+08 | 2.34E+08 | 2.29E+08 |
|  | 4.64E+08 | 2.96E+08 | 2.45E+08 | 1.18E+08 | 2.41E+08 | 1.61E+08 |
|  | 4.43E+08 | 3.20E+08 | 2.93E+08 | 1.98E+08 | 2.27E+08 | 2.43E+08 |
| Average – Concentration / ml | 4.21E+08 | 2.89E+08 | 2.67E+08 | 1.80E+08 | 2.34E+08 | 2.11E+08 |
| Concentration (paticles/frame) | 18 | 12.8 | 13.3 | 11.3 | 11.9 | 11.6 |
|  | 23.5 | 15 | 12.5 | 9.2 | 12.2 | 8.2 |
|  | 22.5 | 16.5 | 14.9 | 10 | 11.5 | 12.3 |
| Average – Concentration / frame | 21.3ea | 14.8ea | 13.6ea | 10.2ea | 11.9ea | 10.7ea |

The trend of the mean value of the passed samples per nanotube length is shown in Fig. 15. The longer the nanotube length, the smaller the mean size of the particles; the mean size of the particles tended to increase as the length of the nanotube decreased (Fig. 15)

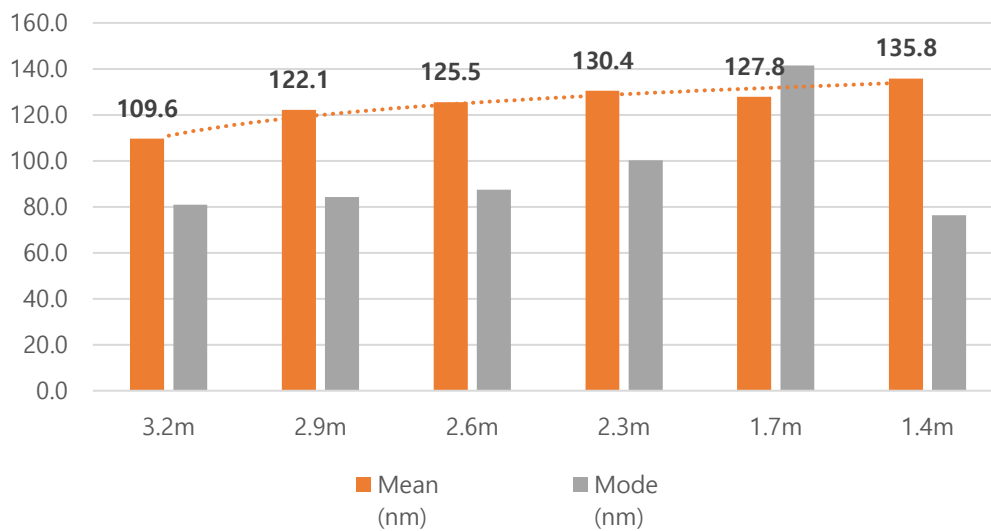

Fig. 15 Particle size per nanotube length – Mean / Mode

The trend of particle concentration of the passed samples per nanotube length is shown in Fig. 16. When the nanotube length was 3.2 meters, the average particle concentration was 4.21e+08 Particles/ml, and when it was 1.4m, the concentration was 2.11e+08 Particles/ml. This result indicates that the shorter the nanotube length, the lower the particle concentration. (Fig. 16)

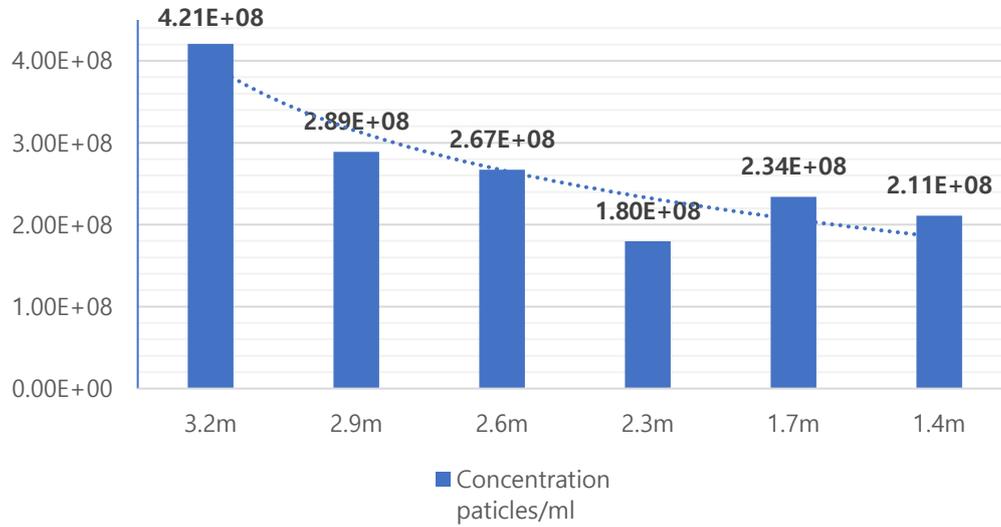

Fig. 16 Particle concentration per nanotube length

The trend of the number of nanoparticles per frame of the passed sample per nanotube length is shown in Fig. 17. When the nanotube length was 3.2 meters, the number of particles was 21.3 Particles/frame, and when it was 1.4m, the number was 10.7 Particles/frame. As such, It was found that the number of particles per frame decreases as the nanotube length decreases, like the concentration of particles (Fig. 17)

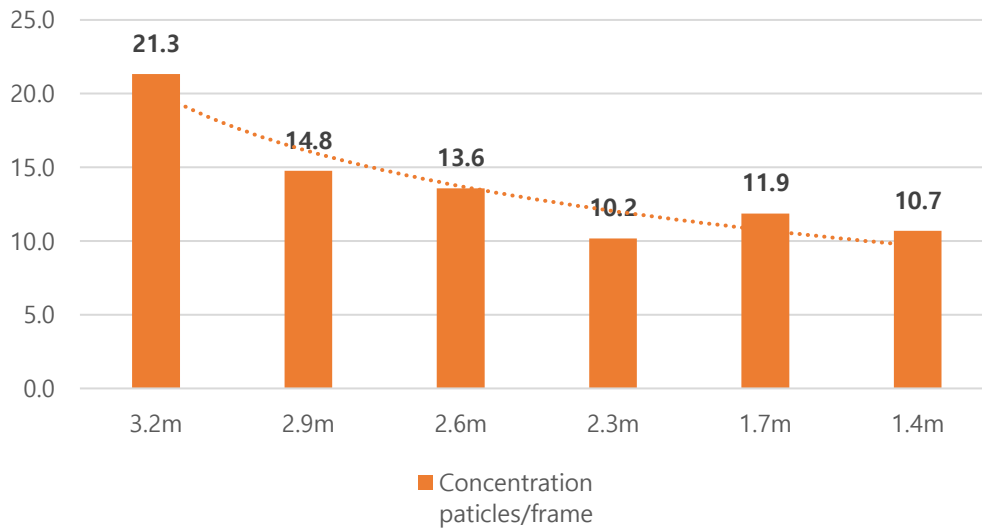

Fig. 17 Number of particles per frame per nanotube length

## 6. Nanobubble Characteristics under Pressure Change

### 6.1 Configuration of Experiment Equipment and Sample Preparation

The size and concentration of particles were confirmed in line with the change in pressure while the nanotube length was fixed. As shown in Fig. 18, a sample tank made of SUS 304 was filled with tap water; a model 'PW-S354M pump for water supply' with a rated output of 350W, a pumping capacity of 26ℓ/min, and a rated power of 220V, 60Hz produced by Wilo was prepared in order to increase the flow velocity while confirming the temperature of 26°C; and the input end of the pump and the output end of the sample tank were connected with a hose.

Two meters of nanotubes with a friction integration of 2.7 and an outer diameter of Ø15 were connected to the output of the pump, and a shearing pressure of 3.5 bar, 2.5 bar, 2 bar, and 1.5 bar was applied using a pressure gauge connected to the inlet of the pump, respectively; nanobubble water was generated by supplying oxygen at 2.0 LPM using an oxygen gas cylinder (bombe) at the pump inlet of the prepared device for generating samples.

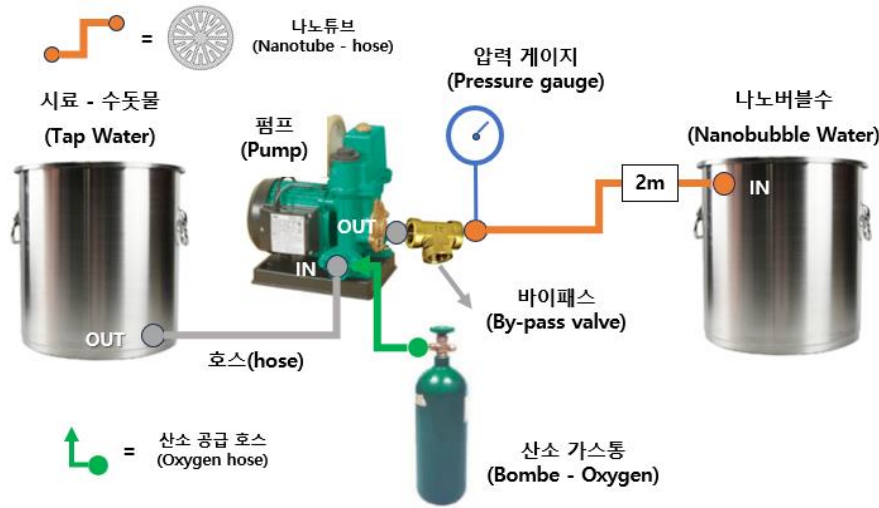

Fig. 18 Diagram of nanobubble generator under pressure change

### 6.2 Experimental Results

As for the size and concentration of the generated nanobubble particles, the same particle analysis tracking device used in Section 3.2 was utilized; to increase the reliability of the particle size analysis results, the same length sample was analyzed three times in a row with an interval of 10 seconds.

The particle size distribution and concentration of the sample passed through the 2-meter nanotube in line with pressure changes are summarized in the table below. As a result of the measurement, the following findings were shown: (1) the average particle size of the sample passed through the nanotube at 3.5 bar pressure was 101.7nm, the particle concentration was 2.27e+08 Particles/ml, and the number of particles in the measurement screen was 11.5 Particles/frame; (2) the average particle size of the sample passed through the nanotube at 2.5 bar pressure was 110.4nm, the particle concentration was 2.19e+08 Particles/ml, and the number of particles in the measurement screen was 11.1 Particles/frame; (3) the average particle size of the sample passed through the nanotube at 2 bar pressure was 125.9nm, the particle concentration was 1.99e+08 Particles/ml, and the number of particles in the measurement screen was 10.1 Particles/frame, (4) the average particle size of the sample passed through the nanotube at 1.5 bar pressure was 121.6nm, the particle concentration was 1.74e+08 Particles/ml, and the number of particles in the measurement screen was 8.8Particles/frame (Table 5)

Table 5. Particle size distribution and concentration measurement results in line with pressure changes in 2m nanotubes

| Water pressure | 3.5 bar | 2.5 bar | 2 bar | 1.5 bar |
|---|---|---|---|---|
| Mean (nm) | 106.4 | 112.7 | 115.2 | 143.5 |
|  | 102.5 | 113.6 | 148 | 113.3 |
|  | 96.3 | 104.9 | 114.5 | 108 |
| Average - Mean | **101.7nm** | **110.4nm** | **125.9nm** | **121.6nm** |
| Mode (nm) | 67.9 | 97.8 | 74.7 | 89.8 |
|  | 82.5 | 90.7 | 84.1 | 108 |
|  | 77.3 | 75.2 | 79.6 | 67.7 |
| Average - Mode | **75.9nm** | **87.9nm** | **79.5nm** | **88.5nm** |
| Concentration (paticles/ml) | 2.21E+08 | 2.09E+08 | 2.09E+08 | 1.55E+08 |
|  | 2.18E+08 | 2.08E+08 | 1.85E+08 | 1.88E+08 |
|  | 2.42E+08 | 2.40E+08 | 2.03E+08 | 1.79E+08 |
| Average – Concentration / ml | **2.27E+08** | **2.19E+08** | **1.99E+08** | **1.74E+08** |
| Concentration (paticles/frame) | 11.2 | 10.6 | 10.6 | 7.9 |
|  | 11.1 | 10.6 | 9.4 | 9.5 |
|  | 12.3 | 12.2 | 10.3 | 9.1 |
| Average – Concentration / frame | **11.5ea** | **11.1ea** | **10.1ea** | **8.8ea** |

The trend of the mean value of the sample passed through the 2m nanotube according to the pressure change is shown in Fig. 19 At the same length, the higher the pressure, the smaller the mean size of the particles; as the pressure decreased, the mean size of the particles tended to increase (Fig. 19)

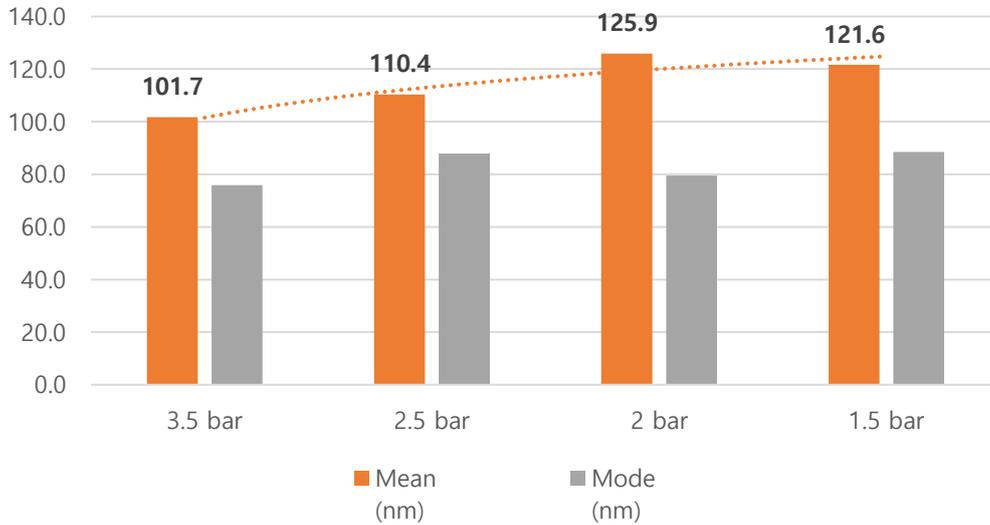

Fig. 19 Particle size in line with pressure changes in 2m nanotubes – Mean / Mode

The particle concentration of the sample passed through the 2m nanotube according to the pressure change is shown in Fig. 20. The mean concentration of the sample passed through the nanotube at 3.5 bar pressure was 2.27e+08 Particles/ml, and at 1.5 bar pressure, the concentration was 1.74e+08 Particles/ml. It was observed that the concentration of particles decreased as the pressure decreased at the same length (Fig. 20)

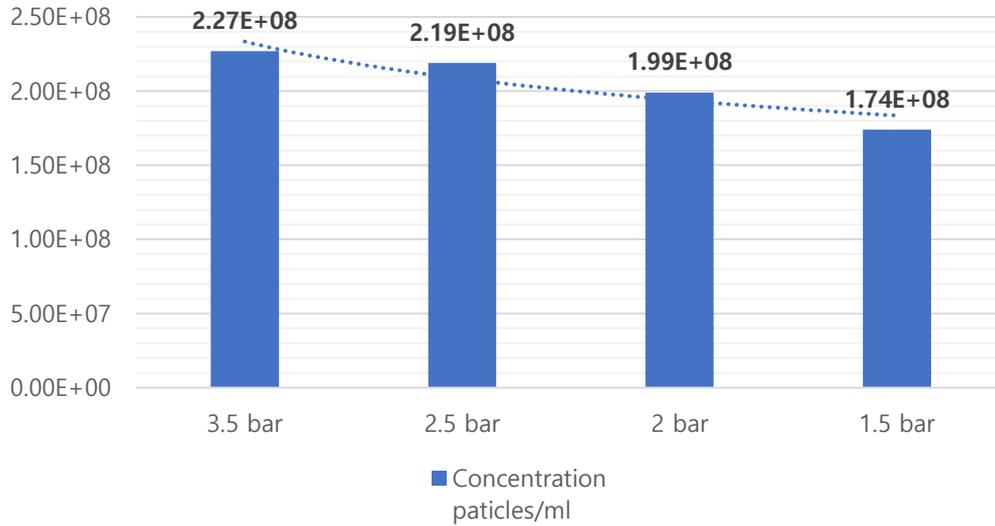

Fig. 20 Particle concentration in line with pressure change in 2m nanotubes

The trend of the number of nanoparticles per frame of the sample passed through the 2m nanotube in line with pressure changes is shown in Fig. 21. The number of particles of the sample passed through the nanotube at 3.5 bar pressure was 11.5 Particles/frame, and at the 1.5 bar pressure, the number was 8.8 Particles/frame. It was found that the number of particles per frame at the same length also decreased as pressure dropped at the same nanotube length (Fig. 21)

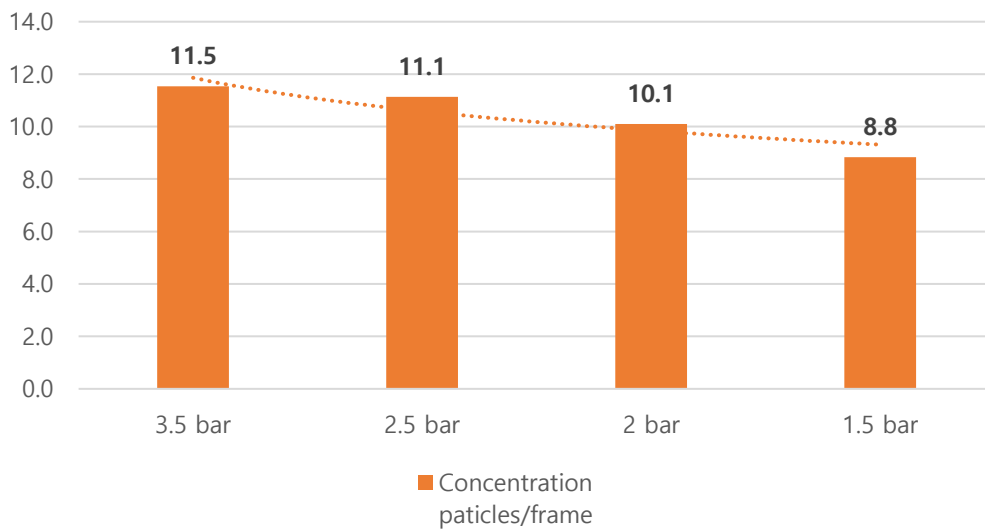

Fig. 21 Number of particles per frame in line with pressure changes in 2m nanotubes

### III. Economic feasibility according to the effective friction constant fluctuations

The higher the friction integration (effective friction constant) and the longer the length, the higher the nanobubble particle concentration, but the lower the overall production. This was confirmed by the results summarized in Table 6. The detailed results are shown in Fig. 22 and Fig. 23. When the effective friction constant was small, the particle concentration became low.

However, when the nanotube composition was varied, there was a difference of 6.28 times in power, approximately 0.43 times in particle concentration, and 41.6 times in daily production. In other words, the efficiency was 2.8 times higher (41.6÷6.28÷0.43=2.8) compared to particle concentration alone.

Table 6. Result comparison in line with composition changes in nanotubes

| Date | 2022. 01. 04 | 2023. 06. 27 |
|---|---|---|
| Type | Type-A | Type-B |
| Friction aggregation | 1.4 | 1.17 |
| Tube length (m) | 5m | 4m(dual) – 2line |
| Effective friction constant | **7** | **4.68** |
| Tube external diameter | Ø25 | Ø90 |
| Pump power | 350W | 2.2kW |
| Inlet pressure | 1.8bar | 1.8bar |
| Daily production | 24 ton | 1,000 ton |
| Concentration (Particles/ml) | **4.87e+08** | **2.13e+08** |
| Mean (nm) | 178.0 | 178.6 |

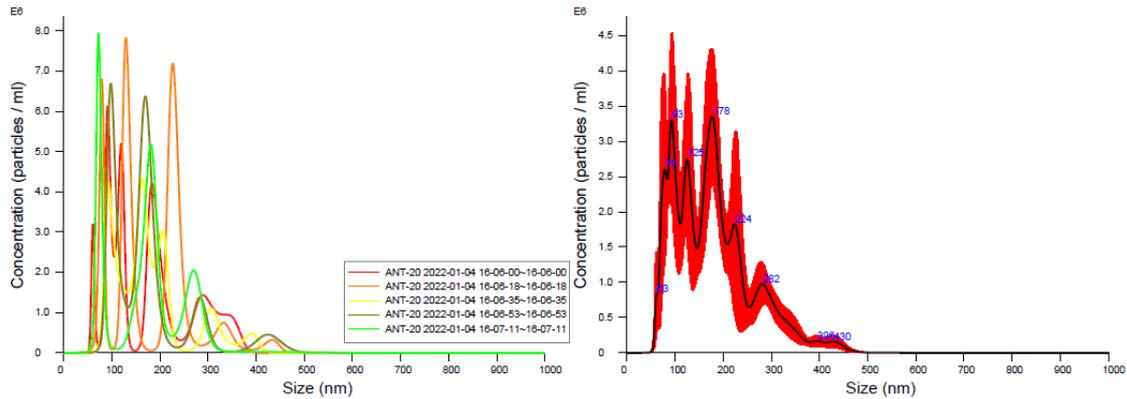

Fig. 22 Measurement of Type-A particle concentration and particle size

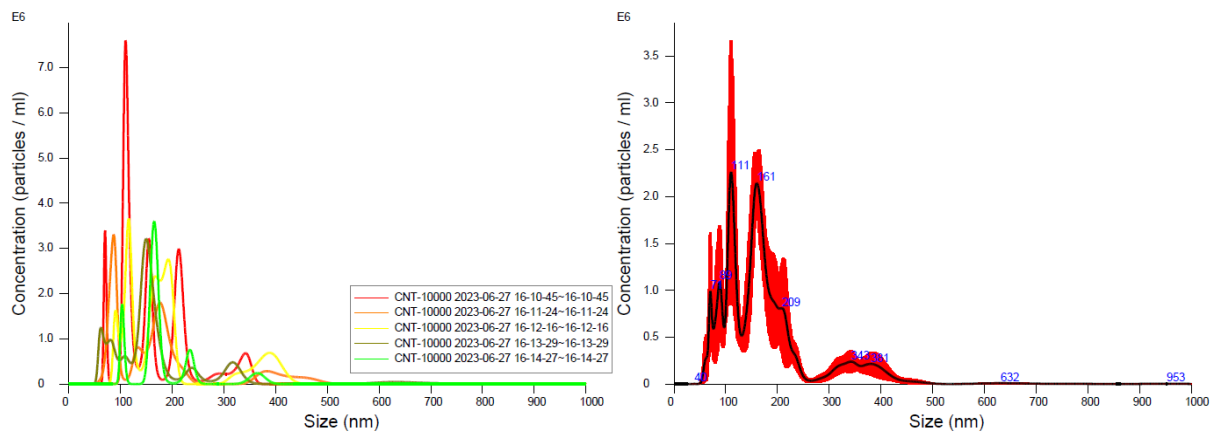

Fig. 23 Measurement of Type-B particle concentration and size

## IV. Conclusion

This study and experiment found that the friction of nanotubes alone can easily and quickly improve the production of nanobubbles with small particle size in real time; by increasing the shearing pressure while increasing the effective friction constant value, the particle size of nanobubbles can be smaller while increasing the particle concentration. The above experiments and measurements confirmed the nanobubble characteristics when the pressure changes were applied in nanotubes of the same length (same effective friction constant). As such, when the bubbles in water move through the nanotubes, they contact the inner surface; here, the bubbles become longer and thinner due to the flow velocity and friction; such longer and thinner bubbles are micronized by repeating the process of splitting; the microbubbles are again nanoized by friction.

In other words, the effective friction constant value can be higher by increasing the friction integration and lengthening the length of the nanotubes, and the flow velocity can be faster by increasing the pressure of the pump, resulting in small particle size and high particle concentration (Fig. 24) the cavitation generation method is not economically feasible because the device is composed of a large-scale heavy body with a complex structure, and the manufacturing cost is excessive, as well as the enormous operation cost from the operation of the rotator. We tried to overcome the disadvantages of the cavitation generation. The nanotubes used in this experiment can significantly shorten the length of the flow path by expanding the friction area per cross-sectional area of the flow path; as the high density of the flow path is possible to be accumulated, the miniaturization and lightweight of the device became possible, as well as the creation of a device with a simple structure and significantly reduced production costs.Additionally, the miniaturization and lightweight of the device using nanotubes can facilitate the treatment and management of the device and improve the spatial utilization, which can greatly expand the range of applications of nanobubbles, not only in large facilities but also in small businesses and households.

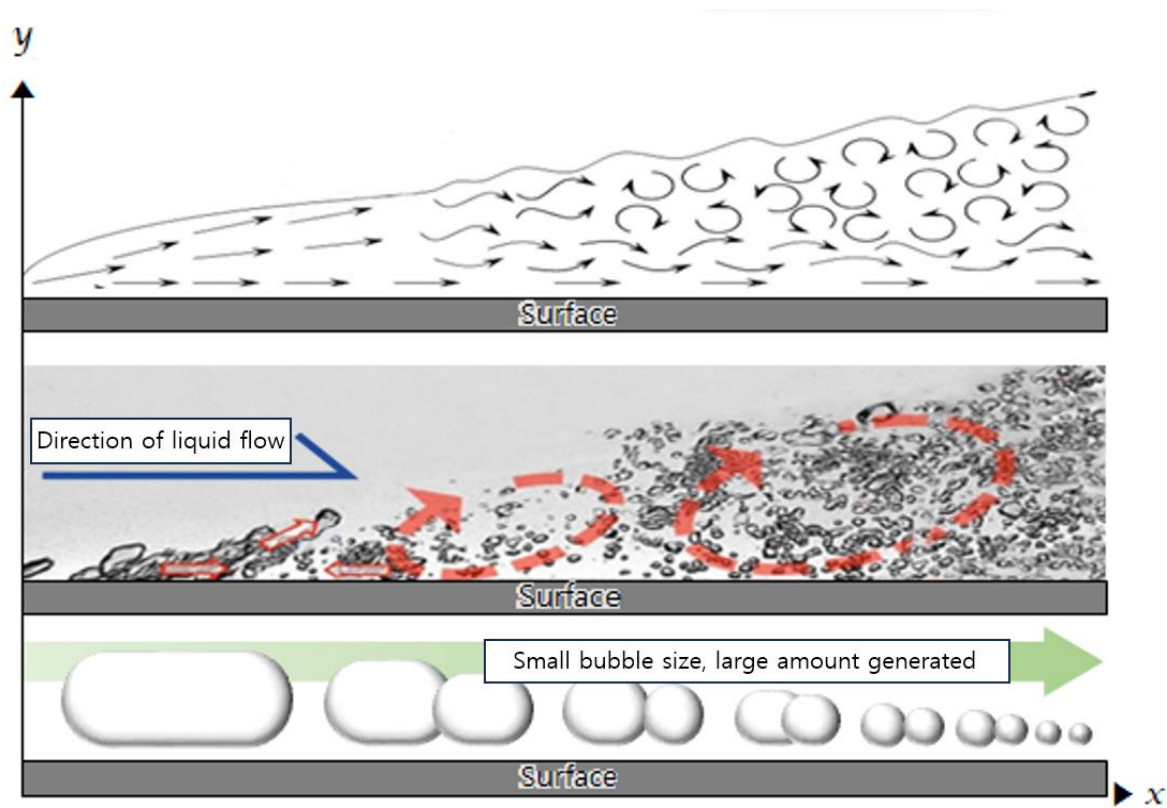

Fig. 24 Mechanism of nanobubble generation by surface friction effect

In addition, as the flow path member for generating microbubbles is made into a tube that can be bent freely, it is expected that it would be facile to install in separate devices requiring various types of bending (e.g., washing machines and home bathtubs), and that it would be possible to dramatically expand the application of nanobubbles in real life and diverse industries.

## References

1. Jia, M. *et al.* Nanobubbles in water and wastewater treatment systems: Small bubbles making big difference. *Water Research* **245**, 120613 (2023).

2. Kyzas, G. Z. & Mitropoulos, A. C. From Bubbles to Nanobubbles. *Nanomaterials* **11**, 2592 (2021).

3. Michailidi, E. D. *et al.* Bulk nanobubbles: Production and investigation of their formation/stability mechanism. *Journal of Colloid and Interface Science* **564**, 371–380 (2020).

4. Nakagawa, M., Kioka, A. & Tagomori, K. Nanobubbles as friction modifier. *Tribology International* **165**, 107333 (2022).

5. Foudas, A. W. *et al.* Fundamentals and applications of nanobubbles: A review. *Chemical Engineering Research and Design* **189**, 64–86 (2023).

6. Kim, J.-Y., Song, M.-G. & Kim, J.-D. Zeta Potential of Nanobubbles Generated by Ultrasonication in Aqueous Alkyl Polyglycoside Solutions. *Journal of Colloid and Interface Science* **223**, 285–291 (2000).

7. Kikuchi, K. *et al.* Concentration determination of oxygen nanobubbles in electrolyzed water. *Journal of Colloid and Interface Science* **329**, 306–309 (2009).

8. V. Postnikov, A., V. Uvarov, I., V. Penkov, N. & B. Svetovoy, V. Collective behavior of bulk nanobubbles produced by alternating polarity electrolysis. *Nanoscale* **10**, 428–435 (2018).

9. Li, T., Cui, Z., Sun, J., Jiang, C. & Li, G. Generation of Bulk Nanobubbles by Self-Developed Venturi-Type Circulation Hydrodynamic Cavitation Device. *Langmuir* (2021) doi:10.1021/acs.langmuir.1c02010.